# Normal incidence narrowband transmission filtering in zero-contrast gratings

Xuan Cui,[1] Hao Tian,[1] Yan Du,[1] Peng Tan,[1] Guang Shi[1] and Zhongxiang Zhou,[1]

[1] *Department of Physics, Harbin Institute of Technology, Harbin 150001, China*

**Abstract:** We report narrowband transmission filtering based on zero-contrast grating (ZCG) reflectors at normal incidence. Computational results show that the filtering is realized through symmetry-protected modes coupling. The guided modes introduced by the slab layer make the filter frequencies flexible to modify. The rectangular structure of the filter allows simple fabrication and integration into optical systems. The quality factor of the filters could exceed $10^6$. Owing to the low refraction index dispersion of the semiconductor and their scale-invariant operations, these filters can be applied in a broad infrared range from near infrared to terahertz wavelengths.

## 1. Introduction

Optical resonators with high quality-factor (Q) modes play crucial roles in modern photonic technologies with applications ranging from sensing, filtering, and display technologies to laser and optical interconnects. The planar design has attracted significant attention in research because of benefits such as easy fabrication and its potential compatibility for on-chip integration with other optoelectronic components. Moreover, dielectric gratings have become attractive planar components for optical engineering, due to their scale-invariant operations in the visible, near-infrared, mid-infrared, and terahertz spectral regions. Recently, high contrast sub-wavelength gratings (HCGs) have demonstrated spectral engineering capabilities, including ultra-broadband reflectors and filters [1–4]. The lateral grating modes, which are called waveguide-array (WGA) modes, result in the extraordinary features of HCG [5, 6]. Furthermore, an improved structure called zero-contrast gratings (ZCG) [7], which solved the discontinuity of structure, has been demenstrated. The additional guided layer introduces the coupling between guided modes and WGA modes, which increases the frequency and angle range of high reflection [8]. Based on HCG reflectiors, through introducing some symmetry-breaking methods, such as oblique incidence and non-rectangle gratings, the symmetry-protected modes (WGA modes) can be coupled to external radiation, resulting in high-Q filtering. Similarly, some kinds of oblique incidence filtrs based on ZCG have been realized. However, normal incidence excitation is desired for many applications, including integrated silicon photonics [9]. Although the HCG filters can work at normal incidence, it is difficult to control the filter frequencies for a traditional HCGs. Moreover, present symmetry breaking methods required either nonequivalent subcells [10] or non-rectangle (rhomboid and right trapezoid) grating bars [11], which significantly complicated the fabrication.

In this Letter, we report narrow band transmission filters based on ZCG at normal incidence. Unlike the traditional guided modes resonance filters, the coupling of the guided modes happens through the symmetry-breaking of a ZCG reflector. The strength of the coupling, as well as the quality factor, can be controlled by the symmetry-breaking level. Furthermore, the filtered frequencies are easily designed through the clear dispersion relationship. The simple structure is much easier to fabricate and integrate into optoelectronic components. Moreover, the filters can be applied in the spectral range from the near-infrared to terahertz regions.

## 2. Guided resonances in a slab

A single slab possesses guided resonances, which are modes that are strongly confined by the slab, but nevertheless can couple to external radiation and therefore in principle have a finite lifetime. The presence of a guided resonance in a slab is manifested as a Fano line shape in the transmission spectrum superimposed on an otherwise smooth background. The resonances can lead to very sharp Fano resonaces if the coupling is weak. Fig. 1(a) shows a schematic of the basic grating structure with the dimensions and incident and transmitted fields defined. The grating dimensions include the period ($\Lambda$), height ($h_1$), and duty cycle ($\eta$), which is defined as the ratio of the high permittivity region ($L$) to the grating period. The frequency ($f$) is normalized by ($c/\Lambda$) in which c is the speed of light in vacuum (in the calculations, we set light speed to unity). For these studies, we utilize transverse magnetic (TM) polarization, which is defined with the magnetic field directed in the $y$-direction. The light is incident from the $z$-direction. The dielectric permittivity ($\varepsilon_r$) is set to 11.9, which is a typical value for silicon in the infrared and terahertz range, and the air permittivity is unity. The structure, shown in Fig. 1(a), is very similar to HCGs except for the duty cycle, and it can be simply considered as a slab layer composed of silicon perturbed by periodic narrow slits.

For HCGs, the lateral modes, or waveguide-array (WGA) modes, of the grating, which were introduced to intuitively explain the high reflectivity of the grating at normal incidence [5], are essential to the transmission properties. Due to the high refractive index of the material, the lateral modes can be maintained in the grating bars despite the subwavelength dimension. The parallel wave vector is introduced by the reciprocal lattice vectors of the grating. In the gratings, the fill factor, namely the air region width, strongly influences the resonance coupling. As the air region becomes ultra-narrow, the coupling is super weak, which leads to high Q-factor Fano resonances.

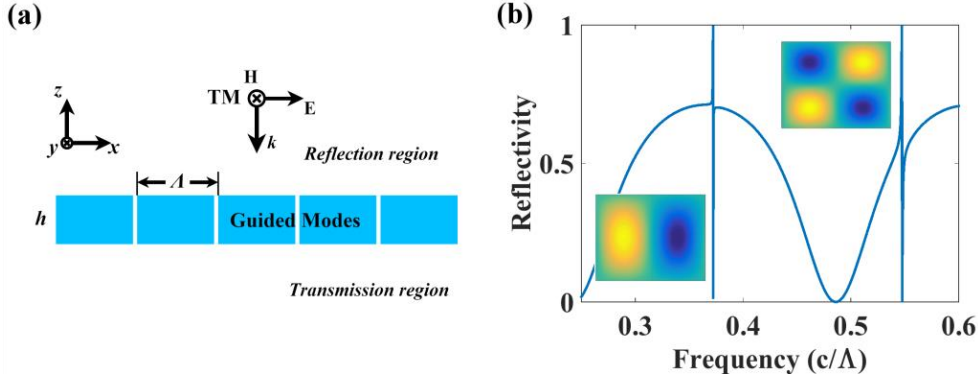

Fig. 1. (a) Grating schematic with the dimensions, incident, and transmitted fields defined. The grating dimensions include the period ($\Lambda$), height ($h_1$), and duty cycle ($\eta$), which is defined as the ratio of the high permittivity ($\varepsilon_d$) region ($L$) to the grating period. The incident light is transverse magnetic (TM) polarization defined with the magnetic field directed in the y-direction. (b) Reflectivity as a function of frequency (c/$\Lambda$) with the parameters $h = 0.6$, $\eta = 0.99$, and $\varepsilon_d = 11.9$. Insets in (b) illustrate the electric field profiles ($E_z$) corresponding to the Fano resonances $TM_1$ and $TM_2$.

When the high index material dominates the grating (Fig. 1(a)), the optical properties of the structure resemble those of a slab waveguide. Guided modes could exist in a slab waveguide. However, these modes do not couple to the radiation modes because of the underlying continuous translational symmetry of the structure [12]. For a grating, the presence of air slits in the slab lowers the translational symmetry of the structure from continuous to discrete symmetry, and thereby some guided modes can couple to radiation modes, which are known as guided resonances [13]. The results of the reflectivity and electric field profile ($E_z$) shown in Fig. 1(b) were computed with the finite-difference time-domain (FDTD) method with the parameters $h_1 = 0.6$ and $\eta = 0.99$ at normal incidence. Two Fano resonances ($f = 0.372$ and $0.548$) superimpose upon the classical Fabry-Perot transmission background. The electric field profiles ($E_z$) are shown at resonance frequencies to illustrate the associated mode coupling. Similar to HCGs, these Fano resonances are caused by lateral bar-mode coupling; however, unlike HCGs, the coupling is weak owing to the extremely narrow air slits in the grating, which provides little transverse momentum for the light wave. In this case, the Fabry-Perot transmission background has not been significantly modulated by these resonances. Therefore, the mechanism of symmetry breaking via narrow slits can be exploited in other grating structures. We will show that this narrow air slit structure can be applied in the ZCG reflector as ultra-high Q-factor ($Q = f/\Delta f$) filters by utilizing the broad opaque band of the ZCG reflector.

## 3. Normal incidence filtering in zero-contrast gratings

ZCG reflectors have a broadband opaque background, and the high reflection range can be optimized by modifying the slab waveguide layer ($h_2$) below the grating, owing to the interference between guided modes and WGA modes. The additional slab layer not only benefits the reflection and fabrication, but also introduces another freedom to engineer the structure. The symmetry of the structure can be broken in both the grating layer and the slab layer. As shown in Fig. 2, for an optimized ZCG reflector, the slits in the slab waveguide layer can realize the coupling between guided modes and external radiation, resulting in filtering, if we put the slits in the assymetric position. The schematic and transmittance are shown in Fig. 2(a) and (b) respectively, with the parameters: $\eta = 0.5$, $h_1 = 0.685\ \Lambda$, and $h_2 = 0.45\ \Lambda$. The slit width($w$) is set to $0.02\ \Lambda$.

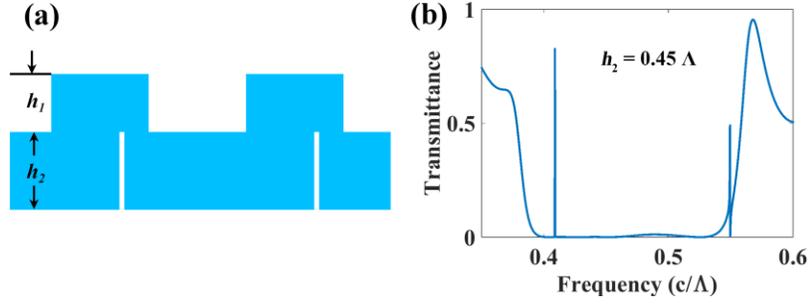

Fig. 2. Grating cross sections with corresponding normal incidence response with the following parameters: $\eta = 0.5$, $h_1 = 0.685\,\Lambda$, $h_2 = 0.45\,\Lambda$, and $w = 0.02\,\Lambda$. The blue areas and white areas in (a) represent the high index material and air, respectively.

In the ZCG reflectors, dark guided resonances that are uncoupled to external radiation exist due to the symmetry protecting. By introducing air slits to the HCG reflectors, the mirror symmetry is broken. When these dark guided resonance frequencies occur in the opaque band, we obtain ZCG filters. For a given structure, the resonance frequencies, namely the filter frequencies, are determined by the internal modes resonances, including WGA modes and guided modes. The high reflectivity of traditional HCG is a result of WGA modes coupling. The coupling of WGA modes is sensitive to the phase changes of the interface. The alteration of the HCG structure might greatly change the reflection. Therefore, it is difficult to control the coupling of WGA modes. In contradistinction to substantial refractive-index discontinuity at the grating/substrate interface for HCG, ZCG refers precisely to the same interface without ambiguity; thereby eliminating local interface reflections and phase changes. Moreover, the guided modes coupling shows a clear dispersion relationship, thus providing the possibility to control the filtered frequencies.

Fig. 3(a) shows the reflectivity contour map versus the normalized frequency ($2\pi c/\Lambda$) and slab-layer thickness $h_2$ for a surface-normal incident TM-plane wave, with the material permittivity $\varepsilon_r = 11.9$ (Si), grating-layer thickness $h_1 = 0.685\Lambda$ and no slits. In this contour map, the different effects of WGA modes and guided modes can be distinguished. A high reflectivity region exists only through the WGA modes coupling at the bottom area of the contour map. In the ZCG, the slab layer shows a guided-modes-like dispersion relationship. As the thickness of slab layer increases, guided modes coupling shows up. When the guided modes possess the similar frequencies of the WGA modes, the range of the high reflectivity is broadened. On the other hand, if the frequencies of guided modes coupling locates in the high reflectivity region, the dispersion lines of guided modes vanish in the contour map. Although the coupling works on the whole structure, through the dispersion of the slab layer it can be treated as guided modes coupling. The first two guided modes introduced by the slab layer are marked by the black dashed lines. These guide modes resonances are uncoupled with the external radiation owing to the symmetry protecting. As shown above, if the asymmetric slits are etched in the slab layer, the guided modes are coupled with the radiation modes, resulting in a Fano resonance. The Fano resonance exhibits nearly perfect transmission as well as perfect reflection over the range of the resonance, assuming the materials are lossless. Consequently, if the resonance occurs in a low transmittance region of the spectrum, the response will yield a transmission filter. Owing to the clear dispersion relationship of the guided modes in the slab layer, we can design the filter frequencies by adapting the slab layer thickness. In this structure, the first two guided modes contribute to the high reflection. Therefore, the filter frequencies are the intersections between the color dashed lines (slab layer thickness) and black dashed lines (guided modes). Fig. 2(a), Fig. 3(b) and Fig. 3(c) are the spectrums at the purple, green and red lines, respectively. In fact, other symmetry-breaking methods can also realize the similar function, such as asymmetric refractive-index distribution, breaking the rectangular symmetry.

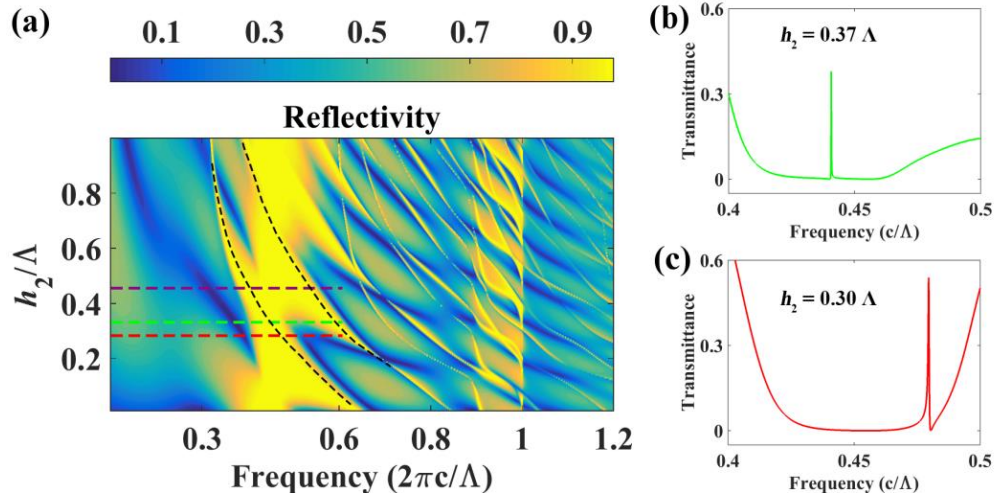

Fig. 3. (a) Reflectivity contour as a function of frequency and slab-layer thickness with $\eta = 0.5$, $h_1 = 0.685\ \Lambda$. The guided modes can be distinguished from WGA modes, which make the filtered frequencies flexible. The transmittance in (b) and (c) represent the slab thickness $h_2 = 0.37\ \Lambda$ (green dashed line in (a)) and $h_2 = 0.30\ \Lambda$ (red dashed line in (a)).

The filtered frequencies are determined by inherent modes of the structure. The quality factor (Q) is mainly determined by the strength of the coupling, namely the slit width in the structure. Fig. 4 illustrates the influence of the slit width ($w$) on the Q-factor, with the parameter: $h_2 = 0.37\ \Lambda$. A similar response is exhibited by the other slab thickness. It is demonstrated that the Q-factor has a negative correlation with the slit width and can exceed a value of $10^6$ at $w = 0.001\ \Lambda$. While optical absorption and experimental constraints limit the attainable Q in practice. Other experimental constraints including the excitation field profile and nonradiative processes can further limit the response.

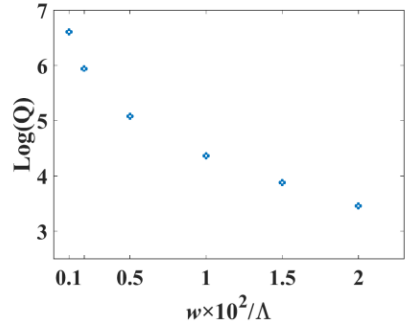

Fig. 4. Filtering quality factor (Q) vs. the slit width.

All simulations are performed with normalized units. Therefore, the structure can be easily designed by adjusting $\Lambda$ for a certain frequency. For example, for the traditional optical communication region, we set the resonance wavelength to 1.55 μm using $h_2 = 0.37\ \Lambda$, and the opaque range is from 1.44 μm to 1.66 μm with the following structure parameters: $\Lambda = 0.683$ μm, $h_1 = 0.468$ μm, $h_2 = 0.308$ μm and $w = 14$ nm. Structures such as those proposed here could be fabricated using advanced nanofabrication techniques [12]. Because the resonance frequencies are insensitive to the position and width of the air slits, the resonance frequencies would not significantly change due to fabrication errors in the ultra-narrow slits. The Q-factor of the structure is larger than $10^4$ for fairly wide slits, so the Q-factor of the filter should be sufficiently high despite fabrication deviations of the slit width.

Here, the permittivity is set to 11.9, which is a typical value of silicon in the infrared range. Silicon and many other semiconductor materials have a high refractive index (2.8–3.5) and little dispersion from the near-infrared to the terahertz region, which enables simple application of the filters in the terahertz region through modification of the grating parameters. The large real and imaginary part of the permittivity of metal in the terahertz region is the primary barrier, causing high losses and the inability to support surface modes. On the other hand, semiconductor materials, which could be treated as dielectrics, can directly apply the optical theory in the terahertz region. Moreover, the fabrication techniques for semiconductor materials are mature. Therefore, the dielectric structure based on semiconductor materials is a promising solution for terahertz devices.

**5. Conclusion**

In conclusion, we proposed a subwavelength grating for realizing high-Q filtering by utilizing the broad opaque band from the ZCG reflector. By introducing narrow slits to the gratings, sharp Fano resonances occur in the original transmission spectrum through symmetry-protected modes coupling. This phenomenon can be applied to some ZCG reflectors for use as high Q-factor filters. These filter structures are flexible to design, simple to fabricate and easy to integrate into optical systems. By optimizing the grating parameters, the Q-factor of the filters reached more than $10^6$. In addition, owing to their scale-invariant operations, these dielectric gratings have promising applications as planar components for optical engineering in the wavelength range from the visible to terahertz spectral regions.